\documentstyle[sprocl]{article}
\def\Journal#1#2#3#4{{#1} {\bf #2}, #3 (#4)}

\def\PRL{\em Phys. Rev. Lett.}
\def\PRB{{\em Phys. Rev.} B}

\def\be{\begin{equation}}
\def\ee{\end{equation}}
\def\bea{\begin{eqnarray}}
\def\eea{\end{eqnarray}}

\begin{document}

\title{QUANTUM HALL EFFECT AT FINITE TEMPERATURES}

\author{ D.G. POLYAKOV }

\address{Institut f\"ur Theoretische Physik der Universit\"at zu
  K\"oln,\\ Z\"ulpicher Str.\ 77, 50937 K\"oln, Germany}

\maketitle\abstracts{Recent work on the temperature-driven
  delocalization in the quantum Hall regime is reviewed, with emphasis
  on the correlation properties of disorder and the role of
  electron-electron interactions.}
 
\section{Introduction}
\baselineskip=11.33pt

The underlying physics of the quantum Hall effect (QHE) is understood
in terms of the Anderson localization in two dimensions (2D): a weakly
disordered 2D electron gas experiences a series of metal-insulator
transitions with increasing magnetic field $B$.\cite{janssen94book}
The field-induced transitions are degenerate in the sense that the
metallic phase occurs only at some particular values of $B$. At zero
temperature, the dissipative conductivity $\sigma_{xx}(B)$ vanishes
everywhere except at a discrete set of the critical points, where it
exhibits peaks of zero width.  My purpose is to summarize recent work
showing how this ``ideal'' picture of QHE evolves with lowering $T$.
Let us first go over some key features of the localization in QHE.

At $T=0$, the crossover to the QHE regime should occur, for weakly
interacting electrons, at $\lambda/l\sim 1$, where $\lambda$ is the
magnetic length, $l$ the mean free path. Yet, making contact with most
of experiments, we are accustomed to treating QHE in the extreme of
high $B$, where disorder-broadened Landau levels are well separated
from each other.  The number of conducting phases, which arise in
succession as the Fermi level sweeps through a single Landau level, is
determined by competition between electron-electron interactions and
disorder.  Since, to date, there exists no reliable theory of
localization in the fractional QHE, I restrict the discussion to the
integer QHE with weak enough interactions. Then there is only one
extended state per Landau level, and its energy $E_c$ coincides with
the center of the level.  Within this framework, the localization
length $\xi(E)$ diverges as $|E-E_c|^{-\gamma}$, where $\gamma\simeq
2.3$.\cite{huckestein95}

Naturally, in a close vicinity of $E_c$, the critical exponent
$\gamma$ should not depend on the correlation radius of disorder, $d$;
however, the range of the universal scaling behavior shrinks if
$\lambda\ll d$.  For the most interesting case of the lowest Landau
level, $\xi(E)$ then scales as
$\xi(\Delta_t)(\Delta_t/|E-E_c|)^\gamma$, only in a narrow band of the
width $\Delta_t=\Gamma(\lambda/d)^2$ around $E_c$, $\Gamma$ being the
width of the Landau level. I refer to this range of $E$ as the
``tunneling band'', for within it the percolating classical
trajectories, which are closed equipotentials in the high-$B$ limit,
are strongly coupled via tunneling through saddle points of the random
potential.  Outside this band, the localization remains classical in
the sense that one may neglect the coupling between the critical
percolating trajectories, the characteristic radius of which behaves
as $R(E)\sim d(\Gamma/|E-E_c|)^{4/3}$.\cite{isichenko92} These
trajectories almost touch each other at the critical saddle points so
as to form the percolation network. However, provided
$|E-E_c|\gg\Delta_t$, the gaps that separate them are typically much
wider than $\lambda$ (to generate the power-law scaling, the critical
saddle points at given $E$ should have energies of the order of
$|E-E_c|$), and so the tunneling plays no role. At the crossover point
between the regimes of the quantum and classical localization,
$\xi(\Delta_t)\sim d(d/\lambda)^{8/3}$. Clearly, this picture rules
out any relation between $\gamma$ (which is close to 7/3) and the
critical exponent of $R(E)$ (exactly equal to 4/3). In the
tunneling-dominated regime, the conducting network is essentially
represented by that of the Chalker-Coddington model\cite{chalker88}
with the characteristic size of the elementary cell $\xi(\Delta_t)$,
independent of $E$. A weak tunneling through chains of non-critical
saddle points does not affect the topology of the network.

\section{Temperature Scaling: Interactions vs Disorder}

In this section, I present an outline of the problem of how
$\sigma_{xx}$ peaks broaden with temperature in the limit of
short-range disorder. Since the localization at $E\to E_c$ originates
from the interference of multiple scattered waves on the large scale
of $\xi(E)$, the problem may be analyzed in terms of the phase
breaking.  The width of the energy band within which the localization
fails to develop, $\Delta_c(T)$, then obeys the simple scaling
relation $\xi(\Delta_c)/L_\phi\sim 1$, where $L_\phi$ is the dephasing
length at the critical energy. The notion of the critical broadening
of the $\sigma_{xx}$ peaks implies that the inelastic scattering is
strong enough, namely $\Delta_c(T)/T\gg 1$.

In trying to apply, after simple modifications, the standard
theory\cite{altshuler85} of the phase breaking in dirty metals to the
metallic phase in the QHE limit, one can immediately see that the
Fermi liquid picture may only be marginally valid. Specifically, it
appears that the crucial parameter $\hbar/T\tau_\phi\sim 1$,
$\tau_\phi$ being the phase-breaking time associated with
electron-electron interactions. In this circumstance, we are only able
to say what is the scaling behavior of $L_\phi$: \be
L_\phi(T)\sim\lambda(\Gamma/T)^{1/2}~. \ee Doing so, however, we run
into a difficulty: for the critical exponent of \mbox{$\Delta_c\propto
T^\kappa$}, Eq.\ (1) gives $\kappa=1/2\gamma$, which is approximately
half of the thoroughly measured, in samples with short-range disorder,
value $\kappa\simeq 0.4$.\cite{wei88} To fix this flaw, we argue that
there exists a crossover temperature $T_c\sim U^2/\Gamma$, below which
the Coulomb interaction between electrons cannot be treated
perturbatively, however weak it is in comparison with disorder. Here
$U$ characterizes the strength of the interaction and scales as
$e^2/\varepsilon\lambda$, $\varepsilon$ being the dielectric
constant. We have to assume the condition $U/\Gamma\ll 1$, in order to
ensure that the interaction does not break down the integer QHE.

We first make the elementary observation that the Maxwell relaxation
of a charged wave packet, built up from eigenstates with energies
close to $E_c$, is characterized, in the quasi-2D geometry, by the
constant velocity of the charge spreading
$v_s=2\pi\sigma_{xx}^p/\varepsilon$.\cite{dyakonov87} Here
$\sigma_{xx}^p\sim e^2/h$ stands for the peak value of the
conductivity in the metallic phase.  It follows that the dynamics of
charged excitations on scales larger than $L_c\sim
D_0/v_s\sim\lambda\Gamma/U$ is no longer diffusive
($D_0\sim\lambda^2\Gamma/\hbar$ is the bare diffusion coefficent,
$L_c$ has the meaning of a screening radius); instead, it is governed
by the interactions. In particular, the spectral function for the
screened density-density correlator in the limit of small $q$ takes
the form \be {\cal S}(\omega,{\bf q})=(\varepsilon q/2\pi
e^2)\times[v_sq/\pi(\omega^2+v_s^2q^2)] \ee [at criticality, the
condition of small $q$ depends on $\omega$: $q$ should be small as
compared to both $L_c^{-1}$ and $\omega/v_s$; if $qv_s/\omega\gg 1$,
the effective velocity $v_s$ in Eq.\ (2) becomes a function of the
ratio $q/\omega$ (a similar renormalization of the effective diffusion
constant at $U=0$ was studied in Ref.\ 8); note also that we deal with
spatial scales smaller than $\xi(E)$ at all $E$ involved (cf.\ Ref.\
9)]. This contrasts with the dynamical properties of non-interacting
electrons, which are controlled, at the critical point, by the
dynamical susceptibility depending on the single parameter
$q^2D_0/\omega$.\cite{chalker882} \mbox{Eq.\ (2)} describes the
Coulomb correlations between charged density fluctuations (originating
from the dynamical screening\cite{altshuler85,siak91,levitov95}). The
relation $\hbar/T\tau_\phi\sim 1$ still holds; however, the Coulomb
correlations should affect the dynamical scaling at criticality,
unless $L_\phi/L_c\ll 1$, i.e.\ $T/T_c\gg 1$. Given that in the
interaction-dominated regime the dynamical scaling can be explained in
terms of the charge-spreading velocity, $L_\phi$ at lower $T$ obeys
the relation \be L_\phi(T)\sim v_s\tau_\phi\sim e^2/\varepsilon T \ee
(borrowing the terminology of thermally driven phase-transitions, one
can say that the dynamical critical exponent $z=1$). This last
expression looks universal since $\sigma_{xx}^p$ does not depend on
the strength of disorder. It is worth noting that the assumption of
the long-range interaction is absolutely crucial in the picture under
discussion. Hence we find that at $T/T_c\ll 1$ \be
\Delta_c\sim\Gamma(T/U)^{1/\gamma}~, \ee which is in decent agreement
with the low-temperature data on samples with short-range
disorder.\cite{polyakov93} In the high-$T$ limit, one should expect a
crossover to the regime of broadening governed by electron-phonon
interactions.\cite{brandes94}

Now for the localization-length exponent $\gamma$, which we keep the
same as for non-interacting electrons. To see how it comes about that
the interaction strongly influences the dynamical behavior of
electrons at criticality, and yet does not change $\gamma$, we recall
that, at $T=0$, even a weak Coulomb interaction between localized
electrons makes the one-particle density of states $g(E)$ vanish at
the Fermi level $E_F$.\cite{shklovskii85} In two dimensions,
$g(E)=c|E-E_F|\varepsilon^2/e^4$, where $c\sim 1$ is a universal
constant. At this point we can already identify $T_c$ with the width
of the gap in $g(E)$ deep in the insulating phase. The concept of the
Coulomb gap was developed for classical electrons, i.e.\ point
charges. It is evident, however, that the same line of argument
applies near the metal-insulator transition as well, only in the range
$|E-E_F|\ll e^2/\varepsilon\xi(E)$. Hence, rather remarkably, $g(E_F)$
vanishes however small $|E_F-E_c|$ is. As a consequence, whatever
$E_F$, the Coulomb energy on the scale of the one-particle
localization length $\xi(E_F)$ is of the order of the characteristic
energy spacing $\delta_c\sim 1/g(E_F+\delta_c)\xi^2(E_F)$ on the same
scale. Within the Coulomb glass approach, this naturally implies that
the long-range interaction merely leads to the repulsion of the levels
of the one-particle states, but it cannot affect the critical behavior
of $\xi(E_F)$.\cite{aleiner93} This conclusion can be reached on more
phenomenological grounds,\cite{lee96} though essentially by a similar
dimension counting. The scaling behavior of $\xi(E_F)$ has been
observed by numerical simulation within the Hartree-Fock
scheme.\cite{ericyang93}. It is worth emphasizing that the range of
energies \be |E-E_F|\ll e^2/\varepsilon\xi(E_F)~, \ee where the above
arguments about the stability of $\xi(E)$ are true, shrinks as $E_F\to
E_c$. Thus the question as to the localization properties of the
excitations at $E_F=E_c$ remains open. Particularly, it is not clear
what kind of the gap in $g(E)$ is observed in the numerical
simulations, when $E_F$ is tuned to be precisely
$E_c$.\cite{ericyang93} It is possible that the gap at $E_F=E_c$ is a
reminiscence of the sharper ``polaronic'' gap, which pertains to clean
systems.\cite{pikus94} Yet, there is no doubt that the true Coulomb
gap survives in the range given by Eq.\ (5), if $E_F\neq E_c$.  We
conclude, therefore, that it is legitimate to exploit the idea of the
one-particle localization at $|E_F-E_c|\sim\Delta_c(T)$, since
$|E-E_F|$ does not exceed $e^2/\varepsilon\xi(\Delta_c)$ within the
range of the temperature smearing of the Fermi distribution.

Thus, near the metal-insulator transition, the notion of the Coulomb
gap matches that of the charge spreading. More specifically, one can
introduce the scale-dependent diffusion coefficient $D(L)\sim
(e^2/h\varepsilon)L$ [see Eq.\ (2)], such that the one-particle
density of states on the scale of $\xi(E_F)$ obeys the relation $\hbar
gD(\xi)\sim 1$. We recognize the latter as a familiar localization
criterion (cf.\ Ref.\ 17). It is worthwhile to notice that, in this
formula, both the diffusion coefficient and the one-particle density
of states are renormalized by the interactions, whereas the Einstein
relation links $\sigma_{xx}^p$ and the bare coefficient $D_0$ via the
thermodynamic density of states. At $T\gg T_c$, the Coulomb gap is
full to the brim and the Coulomb correlations do not control
$L_\phi(T)$ any more (one can then say $z=2$).

The same behavior of $\Delta_c(T)$ is recovered if we use
$L_h(T)/\xi(E)$ as a scaling variable instead of $L_\phi(T)/\xi(E)$,
$L_h\sim [\xi(E)L_\phi]^{1/2}$ [with $L_\phi$ given by Eq.\ (3)] being
the typical hopping length in the insulating
phase.\cite{polyakov93,fisher90} According to this approach, the
conductivity may be represented as
$\sigma_{xx}(x)=(e^2/h)F(x)\exp(-x)$, where $F(x)$ is a dimensionless
power-law function of the single parameter $x=L_h(T)/\xi(E_F)$.  The
advantage here is that the dependence $\ln\sigma_{xx}\sim
-[e^2/\varepsilon\xi(E_F)T]^{1/2}$ can be microscopically grounded in
very general terms for the variable-range hopping regime. It is clear,
however, that this scaling form of $\sigma_{xx}$ explicitly implies
the strong coupling limit, in the sense that it is only valid for
$\hbar/T\tau_\phi \sim 1$.  Otherwise, say if the delocalization would
come from weak interactions with phonons,
$\sigma_{xx}\ll\sigma_{xx}^p$ at $x\sim 1$, which means that there
exists an intermidiate, power-law hopping regime between the metallic
phase and that of the variable-range hopping.

In the critical broadening regime, $\sigma_{xx}^p(T)$ displays only a
weak temperature dependence; however, it might be interesting in its
own right. We first notice that $\sigma_{xx}^p$ is a poorly defined
quantity in coherent samples at zero $T$. Indeed, the conductance
should show strong, sample specific fluctuations of the order of
$e^2/h$ (p.\ 252 in Ref.\ 1). Thus the widely accepted notion that the
conductivity $\sigma_{xx}^p$ at $T\to 0$ has a universal value may
only be relevant if $L_\phi$ is much shorter than the sample size. We
take this limit so as to deal with the self-averaging $\sigma_{xx}^p$.
Then one can identify two contributions to the temperature deviation
$\delta\sigma_{xx}^p(T)$.  One is related to the temperature smearing
of the Fermi distribution and is apparently $\sim
-(e^2/h)[T/\Delta_c(T)]^2$. It is negative and scales as
$T^{2(1-\kappa)}$. The other is similar to the weak-localization
correction, with the difference that the expansion in terms of the
small parameter $\lambda/L_\phi$ should be done around the critical
point. We obtain \be \delta\sigma_{xx}^p\sim
(e^2/h)(\lambda/L_\phi)^{D_2}\propto T^x~,\quad x=\kappa\gamma D_2~,
\ee where $D_2$ may be defined as a generalized dimension of the
critical eigenstates.\cite{huckestein942} For non-interacting
electrons $D_2=2-\eta$, where $\eta\simeq 0.4$ is the critical
exponent of eigenfunction correlations.\cite{chalker882} The problem
is more delicate in the interaction-dominated regime ($T\ll T_c$);
within the framework of the above approach, however, the fractal
dimension $D_2$ in Eq.\ (6) is just half that for non-interacting
electrons. Thus in both cases $x=1-\eta/2$. The scaling arguments
imply that the ``weak delocalization" correction dominates at $T\to
0$. Strictly speaking, near the critical point, the sign of the
correction cannot be obtained within the power-counting analysis.
According to numerical simulations,\cite{huckestein942} however, the
correction in the non-interacting case should be negative for the
lowest Landau level $N=0$. Yet, $\delta\sigma_{xx}^p$ is likely to be
positive for $N\geq 1$ (since $\sigma_{xx}^p$ exceeds the SCBA value
only at $N=0$). Note that $\delta\sigma_{xx}^p$ for higher Landau
levels may be much larger than $e^2/h$, so that at
$\delta\sigma_{xx}^p\sim e^2/h$ a crossover to the logarithmic
dependence\cite{houghton82} on $T$ should occur (in the high-$T$
regime, the ``conventional'' weak-localization correction is
$h\sigma_{xx}^p/e^2$ times smaller than that originating from the
electron-electron interaction).

We might be tempted to conclude that the above concept of the phase
breaking due to the electron-electron interaction completely accounts
for the broadening of $\sigma_{xx}$ peaks at low $T$. In fact, it is
by no means obvious that at finite $T$ the interaction alone is able
to delocalize electrons in the QHE regime. The subtlety is that we
treat the problem of the phase breaking self-consistently, on the
assumption that the inelastic scattering occurs due to electromagnetic
fluctuations produced by delocalized excitations (Nyquist noise). This
approach works perfectly in the weak-localization theory, but in the
QHE it still constitutes a challenging problem because of its
non-perturbative nature.

\section{Long-Range Disorder: Brownian Motion in a Stream}

So far we have dealt with the phase-breaking effects in a short-range
random potential, in which case $\Delta_c(T)/T\gg 1$. As argued in the
introduction, increasing the correlation radius of disorder $d$ brings
the new energy scale $\Delta_t=\Gamma(\lambda/d)^2$ into play. The
ratio $\Delta_c(T)/\Delta_t$ then becomes relevant; specifically,
$\Delta_c(T)$ exhibits the universal scaling behavior in terms of
$\lambda/L_\phi$ only as long as $\Delta_c(T)/\Delta_t\ll 1$. At
$T\ll\Delta_t$, $\Delta_c(T)$ either obeys the relation
$\xi(\Delta_c)/L_\phi\sim 1$ or saturates at the level of
$\Delta_t$. The crucial observation is that the range of $E$ within
which the inelastic scattering prevents the localization,
$|E-E_c|\sim\Delta_c(T)$, cannot exceed $T$ if $\Delta_c(T)$ becomes
wider than the tunneling band. For this reason, the width of the
$\sigma_{xx}$ peak $\Delta\nu(T)\sim\max\{\Delta_c(T),T\}/\Gamma$,
$\nu$ being the filling factor, grows linearly with $T$ at
$T\gg\Delta_t$, irrespective of any particular mechanism of the
inelastic scattering. A sample dependent behavior of $\Delta\nu(T)$,
with a crossover to higher $\kappa$'s as $T$ is increased, indeed was
observed in samples with long-range potential
fluctuations.\cite{koch91}

In the classical regime, the following Fokker-Planck equation typifies
the entire problem: \be {\partial f\over \partial
  t}+{\partial\over\partial\vec\rho}\left[ {\bf v}f-D_i\left({\partial
      f\over\partial\vec\rho}+{\hbar({\bf v}\times{\bf b})\over
      \lambda^2}{\partial f\over\partial E}\right) \right]=0~, \ee
where ${\bf v}(\vec\rho)$ is the drift-velocity field, $\partial {\bf
  v}/\partial\vec\rho=0$, ${\bf b}$ is the unit vector along the
magnetic field. Despite having a simple form, the equation does not
allow for any perturbative treatment of the inelastic scattering
diffusion, characterized by the temperature-dependent coefficient
$D_i(T)$. The point is that, right at the percolation transition, the
critical trajectories are strongly coupled by arbitrarily small $D_i$.
The width of the conducting band then satisfies the self-consistent
equation $\Delta_c\sim\Gamma[D_i/\Omega(\Delta_c)d^2]^{1/2}$,
$\Omega(E)$ being the inverse period of the critical trajectories with
the energy $E$.  The fractal dimensionality of the percolating
trajectories is known to be $7/4$,\cite{isichenko92} which gives
$\Delta_c\sim\Gamma(D_i/D_0)^p$ with $p=3/13$ (in the case of smooth
disorder, $D_0$ can be re-expressed as $D_0\sim <{\bf v}^2>^{1/2}d$).
Accordingly, $\sigma_{xx}^p$ falls off with increasing
$T$ as\cite{polyakov86,isichenko92} \be \sigma_{xx}^p\sim
(e^2/h)(\Gamma/T)[D_i(T)/D_0]^p~. \ee The self-consistent treatment is
necessary only as long as $\Delta_c(T)/T\ll 1$. In the opposite limit,
$1\gg D_i/D_0\gg (T/\Gamma)^{1/p}$, the height of the peak is
determined by the conductivity of the percolation network built up
from trajectories with $|E-E_c|\sim T$; apparently, this yields
$\sigma_{xx}^p\sim e^2/h$.  Interestingly, $D_i$ should be large
enough for this regime to occur, yet neither $\sigma_{xx}^p$ nor
$\Delta\nu$ depends on $D_i$ in this case.  Whatever $D_i/D_0$ is, in
the classical limit $\Delta\nu\sim T/\Gamma$. A noteworthy feature of
the classical delocalization is also the strong enhancement of
$\sigma_{xx}^p$ as compared to the bare value $e^2D_i/\lambda^2 T$.
Notice that the problem can be equivalently formulated in terms of
inhomogeneous local conductivities.\cite{simon94} The above picture of
percolation implies a sharp Fermi distribution for percolating
particles; however, there remains a challenging question: To what
extent is the self-consistent approach adequate in the strongly
correlated electron system with long-range disorder?

\section{Plateau Regime}

In the case of short-range disorder and weak electron-phonon
interactions, one can hardly expect any fascinating features of
$\sigma_{xx}(T)$ deep in the insulating phase at
$|E_F-E_c|\gg\Gamma\gg T$.  Indeed, $\sigma_{xx}$ then
behaves\cite{polyakov94} as $\sigma_0\exp(-|E_F-E_t(T)|/T)$ due to
activation to the energy level $E_t(T)$, such that
$|E_t(T)-E_c|\sim\Gamma\ln^{1/2}(\Gamma/T)$. The pre-exponential
factor $\sigma_0$ is small in comparison with $e^2/h$ and depends on
the electron-phonon coupling constant, though weakly.\cite{polyakov94}
That is why a lively debate arose concerning the universal activated
behavior $\sigma_{xx}=(e^2/h)\exp(-\Delta/T)$, which was reported in a
number of experiments.\cite{clark} The universality of $\sigma_0$
seemed to be intriguing since the activated conductivity occurs solely
due to supposedly weak interactions with phonons. A theory that
explained this feature was forthcoming in terms of the classical
dynamics in a long-range random potential.\cite{polyakov95} If
$d\gg\lambda$, $\sigma_{xx}$ is governed by the activation already at
$|E_F-E_c|\gg T\gg\Delta_t$, once the Fermi level leaves the tunneling
band. Notice, since both the activation exponent and the tunneling one
are linear functions of $E$, the necessary condition of that the
activation dominates is simply $T\gg\Delta_t$. Again, we identify the
crucial parameter $\Omega(T)\tau_T$, where $\tau_T$ is the time it
takes for an electron to change its energy by $T$. The parameter is
small provided $D_i/D_0\gg (T/\Gamma)^{1/p}$. In this limit $\Delta_c$
scales as $T$.  The essential idea of this approach is to represent
the electron system as a random network of thermal reservoirs
connected via ballistic contacts (critical saddle points) with
energies $V_i$ scattered around $E_c$ in the tail of the Fermi
distribution.  This parallels the Landauer-B\"uttiker formalism, with
the difference that the latter was developed for the edge states. The
ballistic conductances of the contacts,
$G_i=(e^2/h)\exp(-|E_c-E_F+V_i|/T)$, exhibit strong asymmetric
fluctuations. However, since $\ln G_i$ are distributed randomly, the
conductivity of the network satisfies the exact relation
$\sigma_{xx}=\exp\left<\ln G_i\right>$ (valid in 2D only), which
immediately gives $\sigma_0=e^2/h$.\cite{polyakov95} If
$\Omega(T)\tau_T\gg 1$, the conducting band $\Delta_c(T)$ gets
narrower than $T$. To put it simply, electrons with $\Delta_c\ll
E-E_c$ are now out of play as they have no time to tune their
electrochemical potential so that it equals the potential of the
adjacent thermal reservoir. Apparently, $\sigma_0=\sigma_{xx}^p$ in
the extreme of high $T$ [see Eq.\ (8)].

\section{Overlapped Levels}

According to the Drude formula, $\sigma_{xx}$ grows with increasing
overlap of Landau levels. For this reason, $\xi(E)$ in the middle
between adjacent $\sigma_{xx}$ peaks acquires an
exponentially large factor, such that $\ln\xi$ scales as
$(h\sigma_{xx}/e^2)^2$. This quickly breaks down QHE at any reasonable
v$T$ (unless the $\sigma_{xx}$ peaks are due to the classical
percolation, which is probably the case in high-mobility samples).
What happens if the number of the coupled levels is limited to
two? Let two Zeeman levels be strongly overlapped in the sense that
$\Gamma\gg\Delta_s$, where $2\Delta_s$ is the difference of the
critical energies $E_c^\pm$ corresponding to two spin
projections. Then turning on a spin-orbit (SO) interaction drives two
systems of electrons with opposite spin into a new
quantum Hall phase with an internal degree of freedom. We present
arguments that the SO coupling is able to greatly facilitate the
inelastic-scattering-induced delocalization.\cite{polyakov952}

In the case of short-range disorder, $\Delta_s$ obeys the
relation\cite{khmelnitskii92} $\xi(E_0)/R_{so}\sim 1$, where
$E_0={1\over 2}(E_c^++E_c^-)$, $R_{so}$ is the spin-flip scattering
length. It follows that, for the weak SO interaction, $\Delta_s$ is
far larger than the local SO splitting. Numerical simulations carried
out in this limit support the conclusion that the critical behavior of
$\xi(E)$ at $E\to E_c^\pm$ remains the same as for spinless
electrons.\cite{lee94} Hence, whatever the ratio
$\Delta_c(T)/\Delta_s$ is, if disorder is short ranged, the SO
coupling does not lead to any principal change in the critical
broadening of the $\sigma_{xx}$ peaks. It merely splits the critical
point [though from an experimental point of view, it might be of great
importance that the SO coupling does yield a sharp numerical growth of
$\xi(E)$]. At $d\gg\lambda$, however, the ``coherent'' contribution to
$\Delta_s$ cannot exceed $\Delta_t$ (the limit $\Delta_s/\Delta_t\sim
1$ then corresponds to the generalized Chalker-Coddington model with
strong spin-flip scattering). In the extreme of smooth disorder,
$\Delta_s$ may be described in completely classical terms as a result
of splitting of percolating classical trajectories.\cite{polyakov96}
Let us formulate an auxiliary percolation problem for particles with
spin: Given two sets of closed equipotentials
$V(\vec\rho)\pm\Delta_s=E$, the particles are allowed to change the
classical trajectories if two equipotentials come within a distance
$\delta$ of each other. The system undergoes a percolation transition
at $\delta=\delta_c(E)$, such that $\delta_c(\pm\Delta_s)$
vanishes. The crucial observation is that $\delta_c(E)\sim
d\Delta_s/\Gamma$ for $|E|<\Delta_s$, but rapidly grows as
$\delta_s\sim d[(|E|-\Delta_s)/\Gamma]^{1/2}$ outside this band, since
at $|E|>\Delta_s$ the percolation occurs only due to the coupling
across the critical saddle points. This simple game leads us to an
interesting result: the strong spin-flip scattering destroys the
classical localization within the energy band
$|E|<\Delta_s$. Consequently, in the limit of long-range disorder
($\Delta_t/\Delta_s\to 0$), when the tunneling through the saddle
points may be neglected, the SO coupling makes $\sigma_{xx}(E_F)$
exhibit a boxlike behavior, namely $\sigma_{xx}\sim e^2/h$ at
$|E_F|\leq\Delta_s$, otherwise $\sigma_{xx}\to 0$. Thus, provided
$T\ll\Delta_s$ and $R(\Delta_s)$ exceeds both $L_\phi$ and $R_{so}$
(the latter was evaluated in Ref.\ 31), two $\sigma_{xx}$ peaks merge
and form the boxlike one. At lower $T$, when $L_\phi$ is still large
in comparison with $R(\Delta_s)$, there are two peaks, but they are
strongly asymmetric. Specifically, $\sigma_{xx}$ between the peaks
falls off with decreasing $T$ only in a power-law manner:
$\sigma_{xx}\sim (e^2/h)R(\Delta_s)/L_\phi(T)$. This power-law hopping
occurs in a wide range of $T$ and goes over into the variable-range
hopping only at very low $T$.\cite{polyakov952} The crucial point is
that the strong spin-flip scattering changes the character of
localization in the energy band limited by $E_c^+$ and $E_c^-$: it
gives rise to the Anderson localization (instead of the classical one)
with $\xi(E_0)/R(\Delta_s)\sim 1$.\cite{polyakov952} We conclude that,
in the case of smooth disorder, the SO coupling is capable of strongly
changing the conventional picture of QHE.

\section{Conclusion} We have discussed inelastic broadening of
$\sigma_{xx}$ peaks in the low-$T$ limit, stressing (i) the crucial
role of electron-electron interactions in the integer QHE; (ii) the
classical aspects of electron dynamics in samples with long-range
disorder.

\section*{Acknowledgments} I am grateful to A. L. Efros, J. Hajdu, and
K. V. Samokhin for interesting discussions.  The work was supported by
the Deutsche Forschungsgemeinschaft.

 \end{document}